\documentstyle[12pt]{article}
\begin{document}
\pagestyle{empty}
\begin{center}
{\bf  FLAVOR MIXING, CP VIOLATION AND A HEAVY TOP }
\footnote
{Invited talk presented at BEYOND THE STANDARD MODEL -IV ,
Lake Tahoe,Tahoe city, California (Dec-13 to Dec-18th,1994)}
\\[.25in]
 Dan Karmgard \& Subhash Rajpoot \\
Department of Physics and Astronomy\\
California State University\\
Long Beach, California 90840\\USA\\[.25in]

\begin{minipage}{4.8in}
{\bf ABSTRACT:} \hbox{} \hspace{.1in}
The recent analysis of the electroweak data at CDF
 indicates that the
constraint on the top quark mass is $m_t=174\pm 17 $GeV. We
accommodate this result in
a new scheme of quark mass matrices in which two elements along the
the  leading diagonal are non-vanishing and the other non-vanishing
elements are those due to  nearest neighbour interactions.
By comparing the quark mixing matrix elements of our scheme with
the experimentally determined Kobayashi- Maskawa matrix elememts in
the standard electroweak model, we find that the CP violating phases
($\delta_1$ and $\delta_2$) in our model violate the CP symmetry
maximally. \\[.15in]
\end{minipage}
\end{center}

\baselineskip 1.5 em

The standard model with three families of quarks and leptons has
eighteen independent parameters
of which ten describe the quark sector.
 The ten parameters are are the six quark masses, the three
 mixing angles and the phase angle responsible for CP violation.
 Eight of the ten parameters are
fairly well known. These are the six quark masses
and the two mixing angles
$Sin{\theta_{12}} $ and $ Sin{\theta_{23}}$ .
The two remaining parameters, $ Sin{\theta_{13}}$, and the CP
 violating phase, $\delta$, are
only poorly determined.
 It is a challenge to
construct a fundamental theory of quark masses and their mixings.
In the absence of such  a theory,
specific form of mass matrices have been put forward in the
literature. The specific form of
quark mass matrices with less than
ten parameters are by now all have been ruled out by the recent
 CDF bound on the mass of the
 top quark, $m_t={174 \pm 17} $  GeV.
We present a specific form of quark mass matrices$^{(1)}$ in which
\baselineskip 1.5 em
there are only two non-vanishing
elements along the diagonal and the other non-vanishing elements
 are those due to nearest
neighbour interactions. In order to limit the number of free parameters,
each mass matrix has is taken to be hermitian. This implies that each
 matrix is thus described
by four real elements and two complex phases. Further, since the
Kobayashi-Maskawa matrix is
a product of two unitary transformations, it is only the relative
 phases that are important.
This limits the
overall number of parameters to just ten.  Explicitly, the mass
 matrices are

\begin{tiny}
\begin{eqnarray}
M_u =\left(
\begin{array}{c c c}
0 & A_u e^{i \alpha_u}& 0\\
A_u e^{-i \alpha_u}& D_u & B_u e^{i \beta_u}\\
0 & B_u e^{-i \beta_u}& C_u
\end{array} \right)
 &  &
M_d =\left(
\begin{array}{c c c}
0  &  A_d e^{i \alpha_d}  &0\\
A_d e^{-i \alpha_d }&  D_d  &  B_d e^{i \beta_d }\\
0  &  B_d e^{-i \beta_d}&  C_d
\end{array} \right)
\end{eqnarray}
\end{tiny}

The matrices $M_u$ and $M_d$ can be expressed in the generic form $M=P {\bf{M}}
P^{\dagger}$
where
\begin{eqnarray}
\bf{M} =\left(
\begin{array}{c c c}
0  &  A  &  0 \\
A  &  D  &  B \\
0  &  B  &  C
\end{array} \right)
 &  &
P= \left(
\begin{array}{c c c}
1  &  0  &  0 \\
0  &  e^{-i\alpha}  &  0 \\
0  &  0  &  e^{-i{(\alpha+\beta)}}
\end{array} \right)
\end{eqnarray}

The eigenvalues $(m_1,-m_2,m_3)$ of $\bf{M}$ are related to the
mass parameters of $\bf{M}$, i.e., $m_1-m_2+m_3=C+D$ ,
$m_1m_2m_3=A^2C$ and
$m_1m_2-m_1m_3-m_2m_3=CD-A^2-B^2 $.

The orthogonal matrix ${\bf O}$
that diagonalises ${\bf M}$ is
 determined to be
\begin{tiny}
\begin{eqnarray}
{\bf O} =\left(
\begin{array}{l l l}
 \sqrt{\frac{(C-m_1)m_2m_3}{C(C-m_2)(C-m_3)}}&
 \sqrt{\frac{(C-m_2)m_1m_3}{C(C-m_1)(C-m_3)}} &
 \sqrt{\frac{(C-m_3)m_1m_2}{C(C-m_1)(C-m_2)}}\\
 \sqrt{-\frac{(C-m_1)m_1}{(m_1-m_3)(m_2+m_1)}} &
 \sqrt{\frac{(C-m_2)m_2}{(m_1+m_2)(m_2+m_3)}} &
 \sqrt{-\frac{(C-m_3)m_3}{(m_3-m_1)(m_2+m_3)}}\\
-\sqrt{\frac{m_1(C-m_2)(C-m_3)}{C(m_2+m_1)(m_1-m_3)}}&
\sqrt{\frac{m_2(C-m_1)C-(m_3)}{C(m_2+m_1)(m_2+m_3)}}&
 -\sqrt{\frac{m_3(C-m_1)(C-m_2)}{C(m_3-m_1)(m_3+m_2)}}
\end{array}
\right)
\end{eqnarray}
\end{tiny}
The matrices ${\bf O_u}$ and ${\bf O_d}$ that diagonalise the
up-type and the down-type quark
 mass matrices $M_u$ and $M_d$ are are gotten by substituting
 $(m_1=m_u,m_2=m_c,m_3=m_t)$
and $(m_1=m_d,m_2=m_s,m_3=m_b)$ in  ${\bf O}$. The
  Cabibbo-Kobayashy-Maskawa matrix $V_{CKM}$
is constructed out
 of the ${\bf O}$'s and the phase matrices $P$'s in the usual
 way, i.e ., $V_{CKM}={\bf O_u}^{T}P_{ud}{\bf O_d}$ ,
$P_{ud}=P_u^{\dagger}P_d $ = diag$(1, e^{-i\delta_1},
e^{-i\delta_2})$, $\delta_1=\alpha_u-\alpha_d$ and
$\delta_2=\alpha_u-\alpha_d+\beta_u-\beta_d$ .
Note that the phase combination $P_u^{\dagger}P_d$ involves
 only the relative phase difference .
Hence there are only two fundamental observable phases in the scheme.

The only free parameters in $ V_{CKM}$ of  our model are $ D^u$,
$D^d$, $\delta_1$, $\delta_2$ as the six quark masses are now known.
We require that the matrix elements of  $ V_{CKM}$ of
our model  lie within one standard deviation of
those determined in  the standard model. This requires
that
$ D^u =1524 \pm 225$ MeV, $D^d =13.5\pm8.5$MeV, $\delta_1=\pi/2$
or $3\pi/2$ while $\delta_2$ is not constrained. The phase angles are
constrained further by comparing the CP violating invariant measure J
of our model with that of the standard model.
The invariant  measure J in the standard model has the constraints,
$5.0$ X 10$^{-5} \leq$ J $\leq$ 1.1 X 10$^{-4}$ for $m_t$ around
180 GeV. For these values of J, the phase angle $\delta$ of the
standard model is constrained to lie between 50$^0$ $\leq$ $ \delta$
$\leq$ 150$^0$. We used  four different combinations of
the matrix elements of $V_{CKM}$ of our model to construct J, i.e,
J=Im$V_{ud}^*V_{us}V_{cd}V_{cs}^*$,
J=Im$V_{cd}^*V_{cs}V_{td}V_{ts}^*$,
J=Im$V_{ud}^*V_{ub}V_{td}V_{tb}^*$,
 J=Im$V_{us}^*V_{ub}V_{cs}V_{cb}^*$
and explicitly verified the invariance of J. We also found that
agreement between the predictions of our model to lie within
  one standard deviation of the experimentally determined central
values constrain the phases
$\delta_1$ and  $\delta_2$ to lie close to $3\pi/2$, a situation that
corresponds to maximal CP violation. Thus our final constraints on
 the four parameters of our model are
$ D^u =1524 \pm 225$ MeV, $D^d =13.5\pm8.5$MeV,
$\delta_1\approx 3\pi/2$
 and $\delta_2\approx 3\pi/2$. It is to be noted that $\delta_1=
\delta_2=3\pi/2$ correspond to an infinite number of solutions for
the phase angles $\alpha_u,\alpha_d,\beta_u, \beta_d$. \\
{\bf ACKNOWLEDGEMENTS:} \hbox{} \hspace{.1in}
This work was
partially supported by  grants from California State University,
Long Beach. We thank Deepak and Jyoti for reading the manuscript.
\begin{center}
{\bf{REFERENCES}}
\end{center}
\begin{enumerate}
\item S. N. Gupta and S. Rajpoot, Quark Mass Matrices and the Top
Quark Mass, Wayne State University Preprint, Phys. Rev. Lett.
manuscript LW4123, L-1 SB/DAE, September 4-th 1990.  S. Rajpoot, Mod.
Phys. Lett. A, Vol. 7, No. 4 (1992) 309
\end{enumerate}
\end{document}